\documentclass[preprint2]{aastex}

\newcommand{\Sp}{{\it Spitzer}}
\newcommand{\etal}{et~al.}

\slugcomment{Submitted to ApJ}

\shorttitle{Clustering of red galaxies around 3C\,270.1}
\shortauthors{M. Haas et al.}

\begin{document}


\title{Clustering of red galaxies around the $z=1.53$ quasar 3C\,270.1}


\author{Martin Haas\altaffilmark{1}, 
  S. P. Willner\altaffilmark{2,3}
  Frank Heymann\altaffilmark{1,4}
  M. L. N. Ashby\altaffilmark{2,3}, 
  G. G. Fazio\altaffilmark{2}, \\
  Belinda J. Wilkes\altaffilmark{2}, 
  Rolf Chini\altaffilmark{1}, 
  Ralf Siebenmorgen\altaffilmark{4},
  Daniel Stern\altaffilmark{5}
}
\altaffiltext{1}{Astronomisches Institut, Ruhr-Universit\"at Bochum,
              Universit\"atsstra{\ss}e 150, 44801 Bochum, Germany,
              email: haas@astro.rub.de}
\altaffiltext{2}{Harvard-Smithsonian Center for Astrophysics, 
  60 Garden Street, Cambridge MA 02138, USA}
\altaffiltext{3}{Visiting Astronomer, MMT, Mt. Hopkins}
\altaffiltext{4}{European Southern Observatory, 
  Karl-Schwarzschildstr. 2, 
             85748 Garching, Germany} 
\altaffiltext{5}{Jet Propulsion Laboratory, 
California Institute of Technology, Pasadena, CA 91109, USA}


\begin{abstract}
  In the paradigm of hierarchical galaxy formation, luminous radio
    galaxies mark mass assembly peaks that should contain 
    clusters of galaxies. 
%    Of particular interest is the epoch when the first clusters of 
%    elliptical galaxies occur.
%    Here we present the first results of a {\it Spitzer Space Telescope}
%    imaging survey of the complete 3CR sample at $1 < z < 2.5$, carried
%    out to characterize these radio sources and their environment.
     Observations  of the $z=1.53$ quasar 3C\,270.1 
     with the {\it Spitzer Space Telescope} at 3.6--24\,$\mu$m and with the
     6.5-m MMT in the  $z'$- and $Y$-bands 
     allow detection of potential cluster members via photometric redshifts.
     Compared with nearby control fields,
     there is an excess of $\sim$11 extremely red objects 
     (EROs) at $1.33\le z_{phot}\le 1.73$, 
     consistent with a proto-cluster around the quasar. 
     The spectral energy distributions (SEDs) of 3/4 of the EROs 
     are better fitted with passive elliptical galaxies than with
     dust-reddened starbursts, and of
     four sources well-detected on an archival HST
     snapshot image, all have undisturbed morphologies.  However,
     one ERO, not covered by the HST image, is a double source with 0\farcs8
     separation on the $z'$ image and a marginal
     (2$\sigma$) 24\,$\mu$m detection indicating a dust-enshrouded
     starburst. 
     The EROs are more luminous than $L^{\star}$
     ($H = -23.6$~AB mag at $z \approx 1.5$).  
%     While spectroscopic corroboration of these results is envisaged, 
%     the first Spitzer study of the environment of a high-redshift 
%     radio-source 
%     already supports the view that galaxy proto-clusters
%     contain massive ellipticals as early as $z=1.5$.
\end{abstract}


\keywords{Galaxies: clustering -- quasars: general -- Infrared: galaxies}



\section{Introduction}
\label{sect_introduction}

Groups and clusters of galaxies are key targets for
studying the evolution of galaxies and large-scale structures.
In the local universe thousands of clusters are known, but 
at high redshift ($z>1$) only few clusters have been identified due to
the challenge of discriminating a distant cluster
from the abundant foreground galaxies.

A few high-redshift clusters have been found via X-ray observations
of a hot intra-cluster medium up to $z=1.45$ (Mullis et al
2005, Bremer \etal\ 2006, Stanford \etal\ 2006, Fang et
al. 2007). Clusters up to this redshift range have also been found  
by searching for galaxy overdensities in 
wide-area optical-infrared surveys, for instance 
the {\it Spitzer}-IRAC shallow survey (Stanford \etal\ 2005, Eisenhardt et
al. 2008),
the UKIDSS ultra-deep survey (van Breukelen \etal\ 2006),
HIROCS/COSMOS (Zatloukal \etal\ 2007),
and the \Sp\ First Look Survey (Muzzin \etal\ 2008).

An alternative way to find high-redshift clusters 
is to use radio sources as signposts. Luminous radio sources occur in
massive galaxies
($>$\,10$^{11}$\,$M_{\odot}$, Seymour \etal\ 2007), 
which according to the hierarchical formation model
represent mass assembly peaks of the early universe. As such, they are
expected to trace overdense regions and therefore cluster formation
sites. The invaluable advantage of searching for clusters around radio
sources is that one knows the redshift to search and therefore one can
limit the observational effort. 

Searches around very-high-redshift ($2<z<5$) radio sources have found
an excess of Lyman\,$\alpha$-emitting galaxies (Kurk \etal\ 2000,
Miley \etal\ 2004, Venemans \etal\ 2007).
These proto-clusters show a rather large
projected extent and only a loose concentration, consistent with
clusters that are forming but not yet virialized
(Intema \etal\ 2006, Zirm \etal\ 2008). 

Based on small ($<$2$\arcmin$) maps, tentative (2--3$\sigma$)
evidence for an overdensity of extremely red objects (EROs, $R-K>5$)
around high-redshift ($1<z<3$) radio sources has been reported
(Best 2000, S\'anchez \& Gonz\'alez-Serrano 2002, 
Stern \etal\ 2003, Toft \etal\ 2003, Wold \etal\ 2003, 
Zirm \etal\ 2008).
These EROs could be high-redshift passive ellipticals or dust-reddened
star-forming galaxies. 
This ambiguity leaves  open the question of 
whether the giant ellipticals were already in place at
these redshifts, implying a formation redshift larger still, or whether the EROs
are starbursts and we are probing the epoch at which the
cluster ellipticals  formed.    

Using the {\it Spitzer Space Telescope} (Werner \etal\ 2004), 
we are performing a 3.6--24\,$\mu$m survey of the
complete 3CR sample of radio galaxies and quasars at $1<z<2.5$ 
(Haas \etal\ 2008).
The large ($>$5$\arcmin$) {\it Spitzer} arrays
enable us to explore also the environment of the
high-redshift radio sources.
The 3.6--24\,$\mu$m wavelength range measures the redshifted stellar
flux peak and
the beginning of the  Rayleigh-Jeans tail,
while dusty starbursts
may in principle show enhanced 24\,$\mu$m emission.
This makes the survey particularly well-suited to identify  
massive ellipticals and dusty starbursts.
This paper presents first results from a pilot study of 
the quasar 3C\,270.1 at $z=1.53$,
combining our {\it Spitzer} data with $z'$- and $Y$-band maps.
Distances and ages in this paper are based on a $\Lambda$CDM cosmology with
$H_0 =71$~km/s/Mpc, $\Omega_{{\rm m}} = 0.27$
and $\Omega_{\Lambda} = 0.73$.

\section{Observations and data}
\label{sect_observations}

%% In a manner similar to \objectname authors can provide links to dataset
%% hosted at participating data centers via the \dataset{} command.  The
%% second curly bracket argument is printed in the text while the first
%% parentheses argument serves as the valid data set identifier.  Large
%% lists of data set are best provided in a table (see Table 3 for an example).
%% Valid data set identifiers should be obtained from the data center that
%% is currently hosting the data.
%%
%% Note that AASTeX interprets everything between the curly braces in the 
%% macro as regular text, so any special characters, e.g. "#" or "_," must be 
%% preceded by a backslash. Otherwise, you will get a LaTeX error when you 
%% compile your manuscript.  Special characters do not 
%% need to be escaped in the optional, square-bracket argument.
 
The {\it Spitzer} maps of 3C\,270.1 were obtained 
using the instruments IRAC (3.6--8.0\,$\mu$m, Fazio \etal\ 2004) and MIPS 
(24\,$\mu$m, Rieke \etal\ 2004). 
On-source exposure times were
4$\times$30\,s (each IRAC band) and 
10$\times$10\,s (MIPS). 
The dithered maps (PID 40072, PI G.\ Fazio) 
cover about 4$\arcmin$ FOV with full depth.
In
addition, comparison fields  offset 6\farcm5 from the main field
were automatically obtained in the IRAC
bands, at 3.6\,$\mu$m and 5.8\,$\mu$m northeast and at 
4.5\,$\mu$m and 8.0\,$\mu$m southwest of the central
field containing the quasar. 
The maps allow us to detect sources down to 
$\sim$4\,$\mu$Jy (3$\sigma$) at 3.6 and 4.5\,$\mu$m
and  100\,$\mu$Jy at 24\,$\mu$m.\footnote{ 
We have also obtained a 16\,$\mu$m map with 
the IRS peak-up array (4$\times$14\,s, Houck \etal\ 2004) which has been
useful for the quasar SED (Haas \etal\ 2008). 
Because of the limited sensitivity (100\,$\mu$Jy) and small size 
(1$\farcm$5$\times$2$\arcmin$), this map does not provide useful 
constraints for the cluster search. 
Therefore, the 16\,$\mu$m data are not further considered here.}

Complementary $z'$- and $Y$-band images were obtained 
at the 6.5-m MMT using MegaCam 
(30$\arcmin$ FOV, Mc\,Leod \etal\ 2006)
and SWIRC (5$\arcmin$ FOV, Brown \etal\ 2008). 
The $z'$- and $Y$-bands 
(centered at 0.89 and 1.05\,$\mu$m, respectively) bracket  
the 4000~\AA~break at redshift $z=1.53$. Sensitivities
better than 1\,$\mu$Jy were  
achieved at $z'$- and $Y$-band under
good seeing conditions (FWHM $<$ 1$\arcsec$) 
with exposure times of 40 and 90~min, respectively. 
The $z'$-band image
covers the IRAC comparison fields, but the $Y$ image does not.
At Galactic latitude $b=80\fdg6$, 
foreground extinction is negligible
($E_{B-V} = 0.012$~mag).

The images were reduced using interactive analysis tools.  
The IRAC mosaics were corrected for the residual images that
arise from prior observations of bright sources by making
object-masked, median-stacked coadds with all the science frames and
then subtracting them from the science frames prior to final coadding.
Thus, we can be sure there are no faint spurious signals in the IRAC
mosaics arising from prior observations.
For IRAC, we used the basic calibrated data products (BCD, version S16) and
coadded them to 0\farcs869 pixels using  version 4.1.2 of
{\tt IRACProc} (Schuster \etal\ 2006). This optimally handles the
slightly undersampled IRAC PSF in order to assure 
accurate point-source photometry. 
For MIPS, we used custom routines to modify
the version S17 BCD files to remove instrumental artifacts (e.g., residual
images) before shifting and coadding to create the final mosaics.
The MegaCam and SWIRC frames were reduced
using standard procedures\footnote{See
http://www.cfa.harvard.edu/$\sim$mashby/megacam/megacam\_frames.html
for MegaCam.}.
%Figure~\ref{fig_map} shows the IRAC 3.6\,$\mu$m image around 3C\,270.1.

The 3.6\,$\mu$m image shows more sources than any other, and we used
it  as the source detection image for constructing a multi-band catalog. 
In the central 4$\arcmin$ $\times$ 4$\arcmin$ with optimal depth,
274 objects are detected at 3.6\,$\mu$m, with 272/202/212 of them 
being also detected at 4.5/5.8/8.0\,$\mu$m At shorter wavelength, 
184 (150) of the 3.6 $\mu$m sources are detected at $z'$
($Y$), and 143 at both $z'$ and $Y$. 
The sources were extracted and matched using the SEXTRACTOR tool
(Bertin \& Arnouts 1996). 
Details are described in Appendix~\ref{appendix_a}. 
The photometric uncertainties are typically less than 10\% but increase
for the faintest sources. 

The region of 3C\,270.1 has also been mapped with the {\it Sloan Digital Sky
Survey} (SDSS, DR6).
About 35 relatively bright SDSS sources are identified
with foreground galaxies
and stars. 
These objects are excluded from the following discussion. 
In addition, we found an archival Hubble Space Telescope WFPC2
snapshot image
of 3C\,270.1 in the F702W passband
(600\,s exposure time, PI W.\ Sparks, Lehnert \etal\ 1999).
Because of the limited spatial coverage and sensitivity (0.6\,$\mu$Jy), 
we did not use the F702W photometry for finding potential cluster
members. However, 
the HST image enables
us to inspect the morphology of a few of the sources to determine
whether they are pointlike or extended  and to look for rest-frame UV
excesses.  

%% In this section, we use  the \subsection command to set off
%% a subsection.  \footnote is used to insert a footnote to the text.


%% This section also includes several of the displayed math environments
%% mentioned in the Author Guide.

\section{Results and discussion}
\label{sect_results}

%% The \notetoeditor{TEXT} command allows the author to communicate
%% information to the copy editor.  This information will appear as a
%% footnote on the printed copy for the manuscript style file.  Nothing will
%% appear on the printed copy if the preprint or
%% preprint2 style files are used.

%% The eqnarray environment produces multi-line display math. The end of
%% each line is marked with a \\. Lines will be numbered unless the \\
%% is preceded by a \nonumber command.
%% Alignment points are marked by ampersands (&). There should be two
%% ampersands (&) per line.

% \begin{figure}
% %\includegraphics[angle=90,scale=.50]{f3.eps}
% \caption{Animation still frame taken from \citet{kim03}.
% This figure is also available as an mpeg
% animation in the electronic edition of the
% {\it Astrophysical Journal}.}
% \end{figure}

% The central 4$\arcmin$\,$\times$\,4$\arcmin$ field 
% around 3C\,270.1 contains 184 galaxies 
% with $z'$ and IRAC photometry. These observed multiband 
% spectral energy distributions (SEDs) are ideally suited to search 
% for intrinsically 
% red objects in the vicinity of the $z = 1.53$ quasar 3C\,270.1. 

\subsection{Cluster galaxy candidates}
\label{sect_candidates}

Cluster galaxies around 3C\,270.1 should lie at the
redshift of the quasar. 
We determined photometric redshifts $z_{phot}$ 
by fitting the observed spectral energy distributions
(SEDs) of the 184 galaxies detected at $z'$ %and 3.6\,$\mu$m 
with two basic templates, an elliptical
galaxy and a dusty starburst galaxy.\footnote{ 
We also tried type-1/-2 AGN templates, but none of the SEDs is
consistent with such templates at redshift $z \sim 1.5$ except
3C\,270.1 itself.} 
For the elliptical galaxy, we used NGC\,221, which has a strong
4000\,\AA~break. 
The NGC\,221 spectral template from Kinney \etal\ (1996), covering the
rest-frame wavelength range 3000-8000\,\AA, 
was smoothly extrapolated to longer wavelengths by a 4000\,K blackbody.
For the starburst galaxy template, we used the ultra-luminous infrared galaxy 
 Arp\,220 with photometry from
the NASA Extragalactic Database (NED) and the {\it Sloan Digital Sky Survey}
(SDSS, DR6) and mid-IR spectra 
from \Sp/IRS.
We also tested the 
dust-reddened star-forming galaxy template M\,82, but the results were
similar to those for Arp\,220 (differences in $z_{phot}$ less than
$\Delta z = 0.1$), 
and therefore we present only the Arp\,220 results.

The accuracy of photometric redshifts can be estimated, 
for instance,  
from the Spitzer Wide-Area Infrared Extragalactic Survey (SWIRE), 
which has spectroscopic redshifts available for many of its sources. 
In SWIRE,  7 filters (5 optical and 2 IRAC) were sufficient to discriminate
between 8 templates (ranging from blue to red galaxy types) 
and to determine $z_{phot}$ with an rms of 
$\Delta z / (1+z) = 3.5\%$ (Rowan-Robinson et al. 2008). 
We have fewer filters than SWIRE, 
but our task is easier because we only have to determine whether the
photometry is consistent with $z=1.53$ or not.  Furthermore, we
consider only extremely red sources and only two templates and
therefore  suggest that we can reach
accuracy similar to that of SWIRE.
At $z=1.5$, an accuracy (rms) of about 4\% corresponds 
to $\Delta z = 0.1$. 
This is much 
larger than the expected redshift dispersion within a cluster 
($\Delta z < 0.01$). 
While the SEDs can provide cluster galaxy candidates within  
appropriate redshift bins around $z_{QSO}$, confirmation of a
cluster will  require spectroscopic redshifts of at least a sample of
the candidates.

\begin{figure}[ht!]
  
  \includegraphics[width=\columnwidth,angle=0,clip=true]{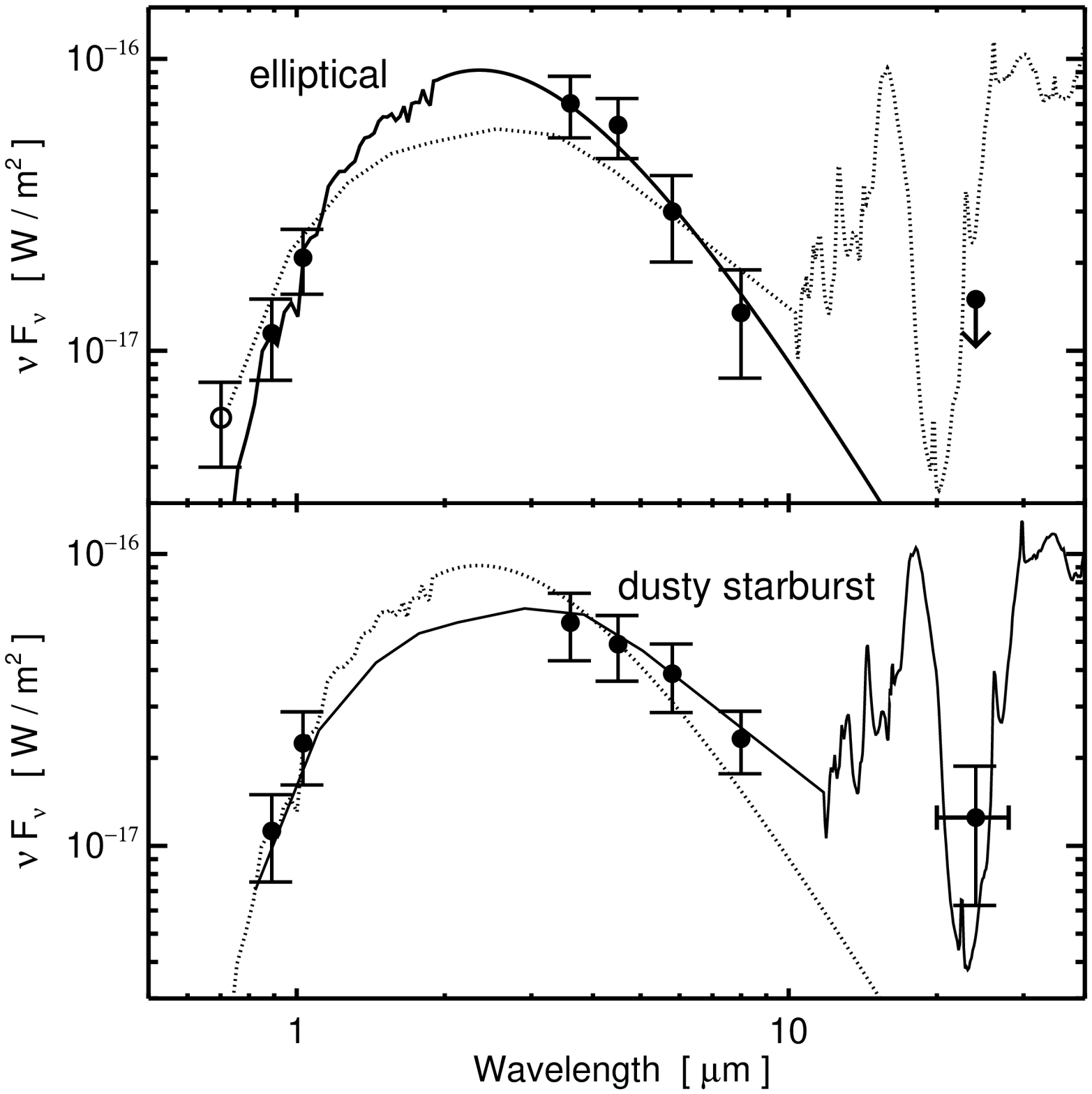}
  
  \caption{Observed spectral energy distributions for two good examples 
    of the 29 cluster candidates. 
    The IRAC and MMT data
    are marked with filled circles 
    and 1$\sigma$ error bars. 
    The MIPS 24\,$\mu$m data point is marked with a filled circle, too;  
    in the upper panel (elliptical source) it is a 3$\sigma$ upper limit, 
    in the lower panel (dusty starburst source) it is a 2$\sigma$
    detection also visible on the map. The horizontal bar indicates the
    24\,$\mu$m pass band for comparison with the silicate absorption 
    feature.
    HST photometry is marked with an open circle; it is 
    not available for the dusty starburst source.  
    Solid lines show the elliptical galaxy (NGC~221) fit for the
    source in the upper panel and the dusty starburst (Arp~220) fit
    for the galaxy in the lower panel.  Dotted lines show the
    alternative template for each galaxy.
    The upper and lower panel shows object numbers 23 and 3,
    respectively, as listed in Table\,\ref{table_1}).
  }
  \label{fig_seds_good}
\end{figure}

\begin{figure}[ht!]
  \includegraphics[width=\columnwidth,angle=0,clip=true]{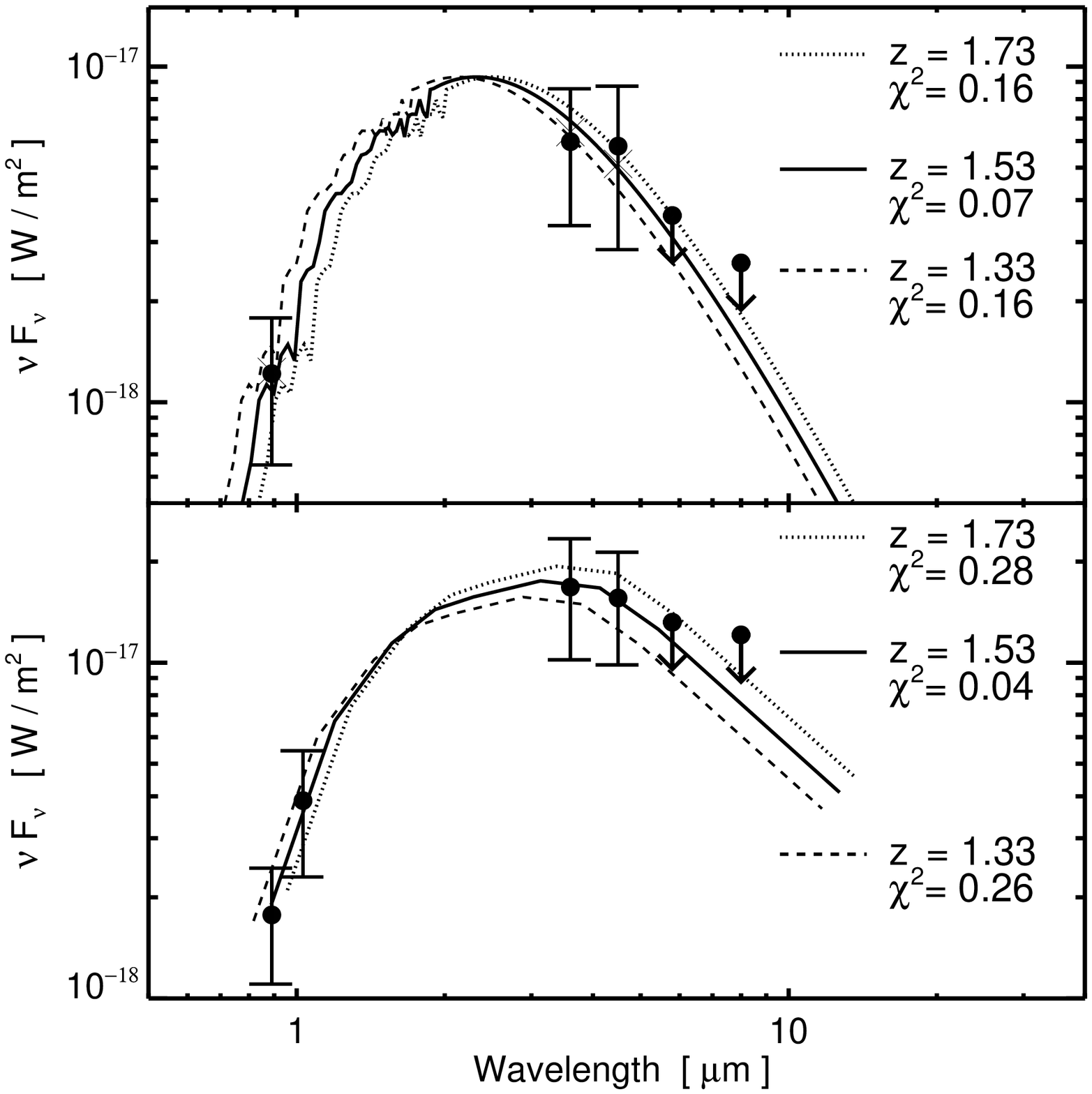}

  \caption{Observed spectral energy distributions for two cluster
    candidates with poorer data quality.  
    The IRAC and MMT data
    are marked with filled circles 
    and 1$\sigma$ error bars. 
    The sources are detected only in the $z'$-band (in the lower panel
    also in the $Y$-band) 
    and at 3.6 and 4.5~$\mu$m, but the upper limits at 
    5.8 and 8.0~$\mu$m help to constrain the redshift fits.
    As in Fig.~\ref{fig_seds_good}, 
    the upper panel shows a source (object 11) which is preferably
    fit by the elliptical template  and the lower panel one (29) fit by
    the starburst template.
    The dotted and dashed lines show the preferred templates  
    at different redshifts (indicated in the figure), showing that
    the accuracy of the photometric redshifts  
    should be  $dz \la 0.2$.
  }
  \label{fig_seds_poor}
\end{figure}

Figures~\ref{fig_seds_good} and \ref{fig_seds_poor} show 
examples of SEDs and template fits to cluster galaxy
candidates. 
The most striking SED features are the steep rise from 
$z'$- to $Y$-band and beyond and the decline between 
3.6 and 8.0\,$\mu$m. 
The slopes of these features determine the redshift. 
We performed the template fits over a grid in redshift 
(dz = 0.01) and intensity. 
For a red SED with $F(3.6)/F(z') \approx 30$, the 
chi square contour plots exhibit a sharp minimum, suggesting that 
for a given template the achievable accuracy of $z_{phot}$ is 
about $\Delta z = 0.1$. 
As illustrated in Fig.~\ref{fig_seds_poor}, even for SEDs with fewer
than six data points the accuracy of $z_{phot}$ should be not worse
than $\Delta z < 0.20$. 
Therefore, 
we have chosen as cluster galaxy candidates the 29 objects for which 
either the elliptical or the starburst template yields 
$1.33 < z_{phot} < 1.73$. 
The basic conclusions on the clustering of EROs around 3C270.1 remain
unchanged for other ranges $|\Delta z|$ between 0.15
and 0.25.

Usually the 0.9--8.0\,$\mu$m SEDs can be fitted very well with both the
elliptical and the starburst template. 
Both templates are extremely red, and
therefore no other galaxy type is likely to fit the SEDs. 
The examples in Figures~\ref{fig_seds_good} and \ref{fig_seds_poor} 
illustrate the ability to identify the two galaxy types at $z \sim 1.53$. 
The resulting fit parameters for the 29 cluster candidates 
are listed in Table~\ref{table_1}.  
The photometric redshifts differ systematically for the 
elliptical and starburst templates. On average,  
$z_{ell}$ is higher by 0.27 $\pm$ 0.1 than $z_{SB}$.  
Among the 29 cluster candidates, the elliptical template 
yields better fits (smaller chi square values) for 22 sources 
(75\%, Table~\ref{table_1}). 
All of these sources lie in the redshift bin $1.33 < z_{SB} < 1.73$.
Among the remaining 7 cluster candidates, 
those 3 sources which favor the starburst fit 
lie in the redshift bin $1.33 < z_{SB} < 1.73$, too. 
These sources have $\chi^2_{\rm SB} < 0.3 \cdot \chi^2_{\rm ell}$  
(object numbers 3, 5 and 29 in Table~\ref{table_1}). 
The remaining 4 sources (object numbers 11, 14, 17 and 28) 
are not detected at 5.8 and 8.0 $\mu$m, and they do not  
clearly favor the starburst template, so that -- 
as ellipticals -- they are cluster galaxy candidates. 

\begin{table*}[t]
\begin{center}
\caption{Fit parameters for the 29 cluster galaxy candidates in the
  quasar field.  
\label{table_1}}
\begin{tabular}{cccccccc}
\tableline\tableline
%Object & RA J2000 & Dec J2000 & $d_{y}$ & $n$ & $\chi^2$ & $R_{maj}$ & $R_{min}$ &
%\multicolumn{1}{c}{$P$\tablenotemark{a}} & $P R_{maj}$ & $P R_{min}$ &
%\multicolumn{1}{c}{$\Theta$\tablenotemark{b}} \\
Object & RA J2000 & Dec J2000 & $n$\tablenotemark{a} &
$z_{ell}$\tablenotemark{b} & $\chi^2_{\rm ell}$ &
$z_{SB}$\tablenotemark{c} & $\chi^2_{\rm SB}$  \\
\tableline
  1 &  12 20 27.90 &  33 43 11.5 &  3 &  1.47 &  0.43 &  1.16 &  0.53 \\ % 235 ell      
  2 &  12 20 28.72 &  33 41 51.9 &  6 &  1.50 &  0.06 &  1.20 &  0.22 \\ % 168 ell
  3 &  12 20 29.20 &  33 42 10.5 &  6 &  1.55 &  0.51 &  1.35 &  0.15 \\ % 176 SB  !?
  4 &  12 20 30.06 &  33 43 18.1 &  6 &  1.47 &  0.19 &  1.00 &  0.85 \\ % 210 ell
  5 &  12 20 30.66 &  33 44 50.0 &  4 &  1.90 &  0.17 &  1.60 &  0.03 \\ % 255 SB  !
  6 &  12 20 30.73 &  33 44 39.4 &  6 &  1.53 &  0.12 &  1.24 &  0.35 \\ % 247 ell
  7 &  12 20 31.18 &  33 43 14.1 &  6 &  1.63 &  0.04 &  1.47 &  0.31 \\ % 190 ell
  8 &  12 20 32.49 &  33 43 20.3 &  5 &  1.50 &  0.29 &  1.34 &  0.39 \\ % 179 ell
  9 &  12 20 33.50 &  33 42 25.4 &  5 &  1.35 &  0.10 &  1.11 &  0.38 \\ % 126 ell HST
 10 &  12 20 33.57 &  33 44 08.3 &  4 &  1.42 &  0.13 &  1.16 &  0.18 \\ % 201 ell
 11 &  12 20 33.85 &  33 45 37.5 &  3 &  1.53 &  0.07 &  1.22 &  0.03 \\ % 252 SB
 12 &  12 20 33.90 &  33 42 34.6 &  4 &  1.36 &  0.08 &  1.17 &  0.30 \\ % 125 ell HST
 13 &  12 20 34.79 &  33 41 55.4 &  5 &  1.47 &  0.10 &  1.07 &  0.27 \\ %  90 ell
 14 &  12 20 34.95 &  33 42 41.0 &  4 &  1.45 &  0.01 &  1.20 &  0.01 \\ % 119 SB or ell HST
 15 &  12 20 35.20 &  33 42 09.6 &  5 &  1.35 &  0.12 &  0.96 &  0.43 \\ %  94 ell HST, chi_arp? SB?
 16 &  12 20 35.88 &  33 43 32.2 &  4 &  1.59 &  0.07 &  1.28 &  0.10 \\ % 141 ell, 3
 17 &  12 20 36.50 &  33 43 30.6 &  4 &  1.40 &  0.19 &  1.20 &  0.10 \\ % 137 SB
 18 &  12 20 37.70 &  33 41 18.2 &  5 &  1.54 &  0.34 &  1.33 &  0.37 \\ %  27 ell
 19 &  12 20 38.52 &  33 43 04.2 &  4 &  1.53 &  0.05 &  1.25 &  0.08 \\ %  88 ell
 20 &  12 20 38.78 &  33 43 50.2 &  6 &  1.68 &  0.32 &  1.38 &  0.53 \\ % 114 ell
 21 &  12 20 39.16 &  33 44 31.9 &  4 &  1.50 &  0.01 &  1.27 &  0.03 \\ % 151 ell, 3
 22 &  12 20 40.10 &  33 43 20.5 &  6 &  1.50 &  0.23 &  1.36 &  0.35 \\ %  83 ell
 23 &  12 20 40.49 &  33 43 21.9 &  6 &  1.56 &  0.13 &  1.04 &  0.98 \\ %  65 ell
 24 &  12 20 40.55 &  33 43 17.3 &  4 &  1.63 &  0.03 &  1.55 &  0.06 \\ %  64 ell, 3
 25 &  12 20 40.62 &  33 43 00.2 &  4 &  1.48 &  0.01 &  1.25 &  0.02 \\ %  53 ell, 3
 26 &  12 20 43.31 &  33 42 51.2 &  5 &  1.50 &  0.44 &  1.32 &  0.66 \\ %  22 ell
 27 &  12 20 43.32 &  33 44 19.8 &  5 &  1.69 &  0.01 &  1.37 &  0.04 \\ %  77 ell, 3
 28 &  12 20 44.17 &  33 44 03.9 &  4 &  1.56 &  0.11 &  1.27 &  0.07 \\ %  51 SB
 29 &  12 20 45.13 &  33 43 18.9 &  4 &  1.84 &  0.26 &  1.53 &  0.04 \\ % 274 SB  !
\tableline
\end{tabular}
%% Any table notes must follow the \end{tabular} command.
%\tablenotetext{a}{Sample footnote for table~\ref{tbl-2} that was
%generated with the \LaTeX\ table environment}
%\tablenotetext{b}{Yet another sample footnote for table~\ref{tbl-2}}
%\tablenotetext{c}{Another sample footnote for table~\ref{tbl-2}}
%\tablecomments{We can also attach a long-ish paragraph of explanatory
%material to a table.}
\tablenotetext{a}{Number of data points (detections) used to calculate
  $\chi^{2}$ of the fit.}
\tablenotetext{b}{Redshift of the fit using the elliptical template NGC\,221}
\tablenotetext{c}{Redshift of the fit using the starburst template Arp\,220}
\end{center}
\end{table*}

\begin{figure}[ht!]
\begin{center}
  \includegraphics[width=\columnwidth,angle=0,clip=true]{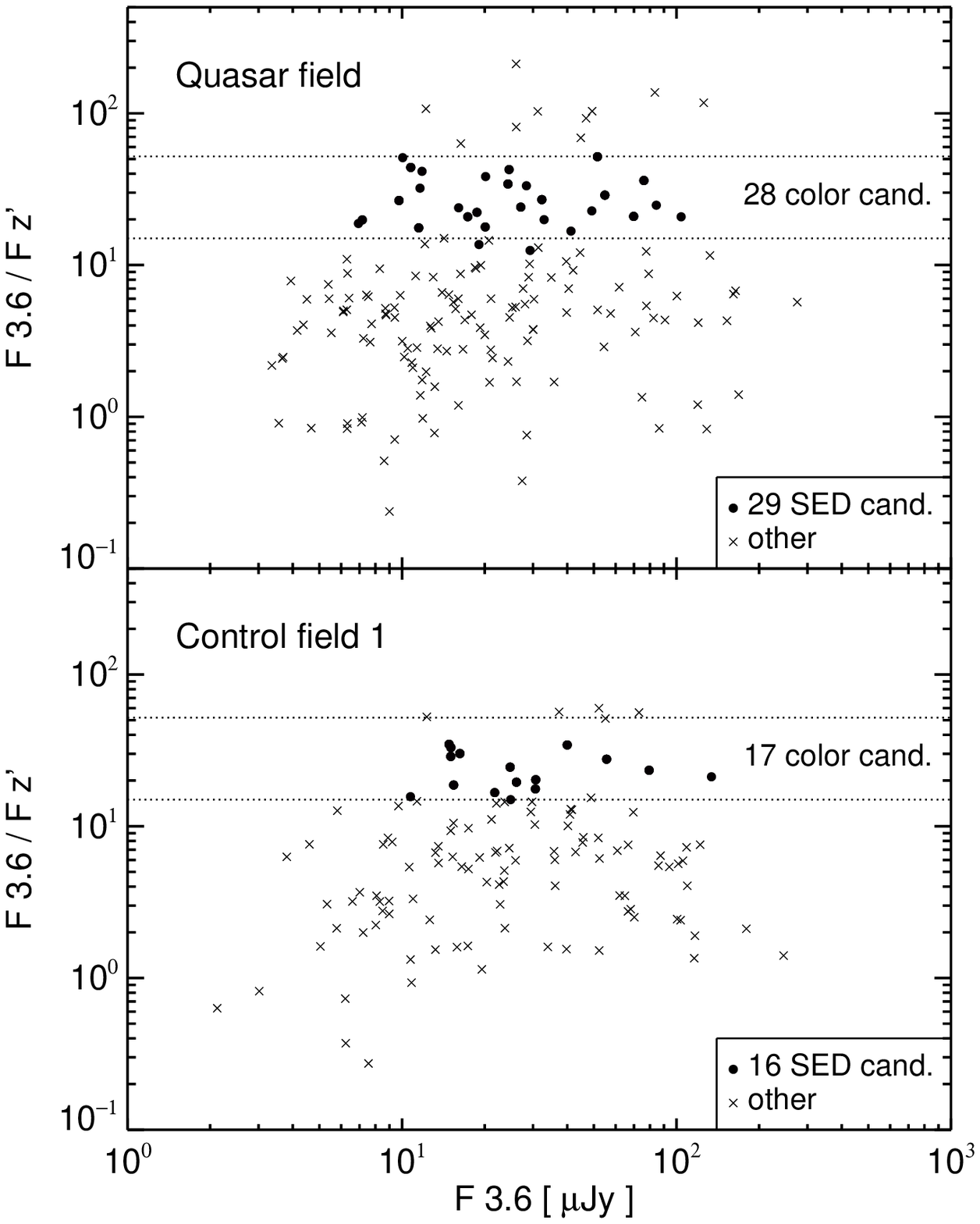}

  \caption{Color-magnitude diagram $F(3.6) / F(z')$ versus  $F(3.6)$,
    for the quasar field (upper panel) and control field 1 (lower panel). 
    Sources with $z_{phot} = 1.53 \pm 0.20$ (SED determined candidates marked with
    circles) concentrate in
    a distinct color range $15 < F(3.6) / F(z') < 52$ indicated by
    the horizontal dotted lines. 
    The number of color determined candidates is also given.
  }
  \label{fig_cm}
\end{center}
\end{figure}

The cluster galaxy candidates occupy a limited range of $z' - [3.6]$
color as shown in Figure \ref{fig_cm}. 
While the candidates were determined from multi-band SED
fitting (usually 4--6 bands), 
selecting a color range  of 
$15 < F(3.6)/F(z') < 52$ would have given nearly the  same sample.
The sources redder than this color range are probably 
at larger redshift ($z>1.8$),
while bluer sources 
are consistent with being either foreground objects or unreddened
star-forming galaxies at the quasar redshift.
Our EROs with $F(3.6) / F(z') > 15$ (corresponding to
$z'-[3.6]>5$\,mag in the Vega system) are likely to also obey the standard
ERO definition of $R-K>5$\,mag
(cf.\ Wilson \etal\ 2004). 

\subsection{Sky distribution and surface density}
\label{sect_sky}

\begin{figure}[ht!]
  \includegraphics[height=\columnwidth,angle=90,clip=true]{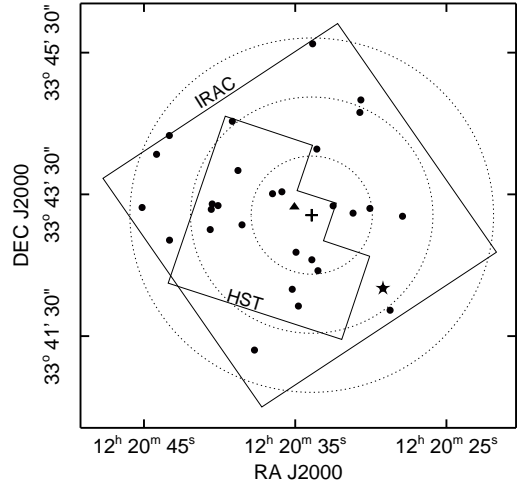}
  \caption{Sky distribution of the 29 candidate cluster galaxies around the
    quasar 3C\,270.1 (marked with a cross). 
    The triangle shows the apparent centroid of the 
    cluster galaxy distribution.
    The star in the southwest
    marks the starburst candidate whose SED is shown in
    Fig.~\ref{fig_seds_good}.   
    The solid lines surround the areas covered by IRAC and HST frames.
    The dotted circles of radius 50\arcsec,
    100\arcsec, and 150\arcsec\ outline the areas considered 
    in Fig.\,\ref{fig_radius}, they are centred around the quasar. 
    At $z=1.53$, 50\arcsec\ corresponds to 427 comoving kpc.
  }
  \label{fig_sky}
\end{figure}

\begin{figure}[ht!]
  \includegraphics[height=\columnwidth,angle=90,clip=true]{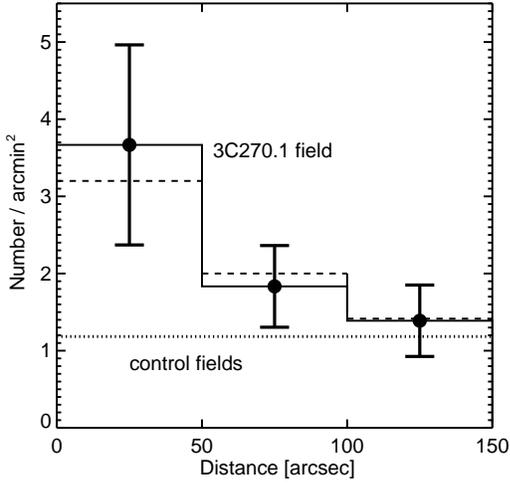}
  \caption{Surface density of the 29 cluster galaxy candidates 
    versus projected distance from  the quasar
    3C270.1 (solid line through fat dots with Poisson error bars).
    The radial bins centred around the quasar 
    are outlined in Fig.~\ref{fig_sky}, and the surface
    density of the outermost annulus has been corrected for the area not
    covered by IRAC. For comparison, the long-dashed histogram shows
    the surface density of cluster galaxy candidates 
    versus projected distance from centroid (the triangle in
    Fig.~\ref{fig_sky}). 
    The dotted line indicates the mean surface density in the
    two control fields.
  }
  \label{fig_radius}
\end{figure}

Figure \ref{fig_sky} shows the sky distribution of the cluster candidates. 
Spreading over a diameter of more than 2\arcmin, they form at most a loose 
concentration around the quasar, indicating a
proto-cluster rather than a virialised, concentrated system.
The peak density appears to lie $\sim$30$\arcsec$ to the east of
3C\,270.1, but it is difficult to be precise with so few galaxies.
%Their centroid is overplotted as filled triangle in Fig. \ref{fig_sky}. 
The 3C\,270.1 proto-cluster is larger in angular size than the X-ray
clusters at $z=1.45$ and $z=1.22$ found by Mullis \etal\ (2005) and
Bremer \etal\ (2006). These clusters 
have an extent of less than 1\arcmin\ in both their galaxy
distributions and their X-ray sizes.

We have also analyzed the two comparison fields taken with
IRAC and covered by the MegaCam $z'$ image (but not the SWIRC $Y$
image). Details of the analysis are described in Appendix~\ref{appendix_b}.
The surface density of possible $z=1.53 \pm 0.20$ galaxies is
less than about 18 objects in each control field of area
15.2~arcmin$^{2}$ (1.2 per arcmin$^2$). 
Thus, in the redshift range $z = 1.53 \pm 0.20$,
the central field surrounding the quasar shows an excess of 
at least about $29 - 18 = 11$
sources, i.e., 60\% over the comparison fields. 

Figure \ref{fig_radius} shows the radial surface density plot of the
29 EROs in the redshift range $z = 1.53 \pm 0.20$. 
The surface density peaks inside the central 50\arcsec\ radius and 
declines steadily with increasing 
distance down to the surface density of the control fields.
This provides further evidence that there is an excess of
$z_{phot}\approx 1.53$ galaxies near 3C\,270.1.
%The overdensity would be even larger if we had used circles centered
%to the east of the quasar position. 
The radial overdensity is also present when using circles around 
the centroid east of the quasar position. 

Existing X-ray data on 3C\,270.1 are inconclusive about the presence
of a cluster.  The quasar itself is clearly detected as a point
source in our 10~ks Chandra data (Wilkes et al., in preparation).  
Weak extended X-ray emission
($<$20\arcsec) is also present, but most of the counts come from the
position of the southern radio lobe or from between the northern
radio lobe and the quasar.  ROSAT data (19.3~ks, 1993 May) show a
strong detection of 3C\,270.1, but at ROSAT resolution not only is
there no way to separate cluster gas emission from quasar emission,
but the emission is also heavily blended with that of unrelated QSO
B1223+338B located 0\farcm8 away.\footnote{
Because of its lower redshift, $z=1.038$, this
QSO does not affect the clustering evidence discussed here.}
We found no XMM observations.
Longer Chandra exposures will be needed for a definitive detection of
any X-ray emission from cluster gas.

\subsection{Nature of cluster galaxy candidates}
\label{sect_nature}

The cluster galaxy candidates have an absolute rest-frame 
magnitude $H \approx K < -23.8$ (AB system). 
An $L^{\star}$ galaxy at $z \sim 1.5$ has $H \approx K =  -23.6$ 
(as determined from observations at $1<z<1.3$ by De~Propris \etal\ 2007). 
Thus only the most luminous cluster galaxy candidates are detected on our maps.
The galaxies are either giant ellipticals 
or dust-reddened starbursts. 
If giant ellipticals, most of their stellar mass has been formed at even higher
redshift, but they may harbor
some ongoing star-formation --- it would be 
relatively weak with respect to the already existing stellar mass 
but could show up at rest UV wavelengths. 
If dusty starbursts, they would have to be at least three
times more luminous in the near-infrared than the ULIRG Arp\,220,
indicating that a large stellar mass has already formed during an
earlier episode.

The distinction between ellipticals and dusty starbursts is difficult
to make at rest wavelengths shorter than 1\,$\mu$m using photometry only, 
albeit possible with adequate wavelength coverage 
(Pozzetti \& Mannucci 2000).  At $z = 1.5$, the six data points of  
our SEDs cover about rest wavelength 0.3--3\,$\mu$m.
As mentioned in Sect.\,\ref{sect_candidates}, 
75\% of the SEDs are better fitted by the
elliptical than the starburst template. 
In principle, additional sensitive photometry  
at rest wavelengths longer than 10\,$\mu$m could improve 
the starburst--elliptical distinction (Stern \etal\ 2006). 
At $z=1.53$, unfortunately, the 9.7$\mu$m silicate absorption enters the 
24\,$\mu$m band, reducing the potential to detect the powerful MIR
emission of starbursts. 
Apart from 3C\,270.1 itself,
there is only one cluster candidate with a marginal (2$\sigma$) detection
at 24\,$\mu$m (Fig.\,\ref{fig_seds_poor}). Its 
SED provides evidence for a dust-enshrouded starburst. 
While only elongated on the $Y$-band image, on the $z'$-band image
this galaxy appears as a double source
with 0\farcs8 ($\sim$7\,kpc) separation; for determining $z_{phot}$
we have used the combined $z'$-band photometry. This source is not
covered by the HST image (Fig.~\ref{fig_sky}).
 
\begin{figure}[ht!]
    \includegraphics[width=\columnwidth,angle=0,clip=true]{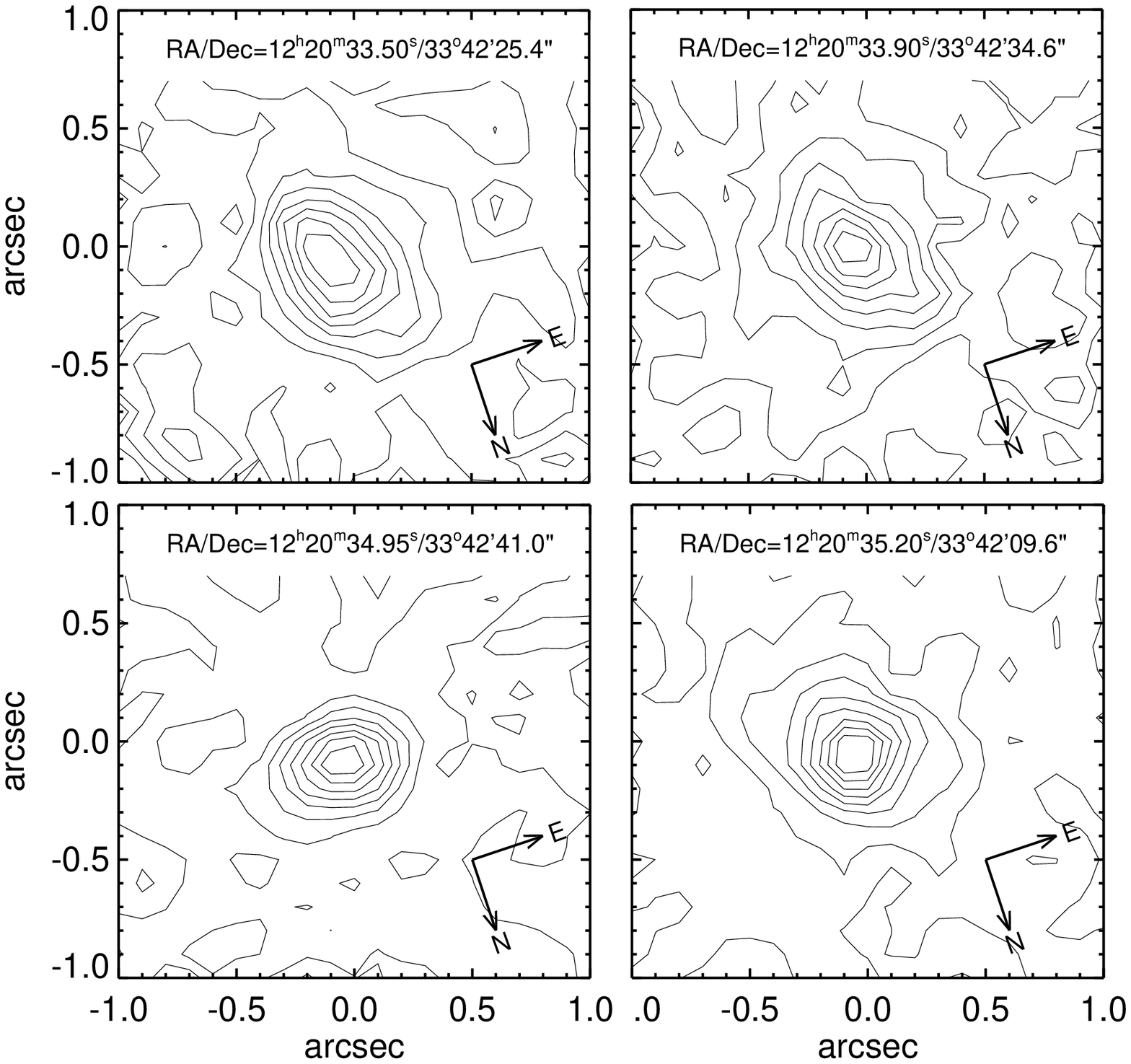}
    \caption{HST F702W images of four cluster galaxy candidates
      (object numbers 9, 12, 14 and 15 in Table~\ref{table_1}). 
      The contours are linearly spaced in steps of 10\% of the peak
      flux value (0\%, 10\%, ... 90\%).  Arrows indicate the
      orientation of each panel.
    }
    \label{fig_hubble_stamps} 
\end{figure}

Because the $z'$-band image is limited in spatial resolution, 
we have also inspected the morphology of the sources seen on the HST
snapshot image.  
The HST frame covers 15 potential cluster galaxies (Fig.~\ref{fig_sky}), 
9 of which are detected.  All 9 are extended, ruling out the (remote)
possibility that they might have been brown dwarfs.
Figure~\ref{fig_hubble_stamps} shows the four sources 
%(object numbers 9, 12, 14 and 15 in Table~\ref{table_1}) 
detected in the 
F702W filter with photometric accuracy better than 5$\sigma$.
The SEDs of these galaxies are best fit with the elliptical template.
They have a regular shape, and none of them shows a peculiar morphology.
This argues against starburst galaxy pairs and in favor of 
an evolved elliptical population. 
The remaining five of the 9 HST-detected sources  are too faint to
draw stringent morphological conclusions. 

The HST F702W photometry is of limited use because of the short
exposure time, but six of the nine detected sources appear to show 
a rest-frame UV excess above the elliptical template
(Fig.~\ref{fig_seds_poor}).
This indicates some level of ongoing star formation activity.
However, at rest wavelengths around 1\,$\mu$m this activity 
appears to be outshone by large numbers of already-evolved stars.
%Our present data do not allow us to discriminate between various
%models inferring the star formation history of the ellipticals
%(Bruzual \& Charlot 2003, Maraston \etal\ 2005).  

Having so many as $\sim$11 
galaxies brighter than $L^{\star}$ would be very high for a local
cluster, but passive evolution will cause these galaxies to become
fainter by something like 3~mag\footnote{
Passive evolution was estimated via a Starburst99 model
(V{\'a}zquez \& Leitherer 2005) with an instantaneous burst age 1~Gyr
before $z=1.53$ and thus an age of 10~Gyr at $z=0$.}
by $z=0$ to become comparable to today's cluster ellipticals. 
Thus the photometric data are compatible with the existence of a cluster.

\section{Conclusions}
\label{sect_conclusions}

High redshift radio sources are predicted to serve as signposts 
for galaxy clusters in the early universe. 
In a pilot study using sensitive, large maps from the 
{\it Spitzer Space Telescope} and the 6.5-m MMT,  
we find a clustering of 
red galaxies around the $z=1.53$ quasar 3C\,270.1:

\begin{itemize}

\item[1)] Photometric redshifts identified 29 galaxies consistent
with the quasar's redshift.
While moderate clustering is evident in both the radial surface
density of these objects and in their numbers
with respect to nearby control fields, 
the sky distribution of the EROs suggests a loose, 
not-yet-virialized protocluster.  
Deep, high-angular-resolution X-ray observations are required to see whether
hot intra-cluster gas is present or not.

\item[2)] The cluster-candidate galaxies have $z'-[3.6]$ colors
indicative of extremely red objects (EROs).  They may be either
dusty starbursts or elliptical galaxies with little ongoing star
formation. The SED fits  
favor the majority of the potential cluster members being 
passive ellipticals, and this is consistent with the four objects
with good HST images having undisturbed morphology.
In some cases the HST photometry reveals a rest-frame UV excess over
the elliptical template, consistent with modest on-going 
star formation and/or nuclear activity in these galaxies, but they
must have formed the bulk of their stars
at redshifts $>$1.53.
 
\end{itemize}

This pilot study demonstrates that the 
{\it Spitzer}/ IRAC maps provide an efficient way to search
for clustering of red galaxies around high redshift radio sources, but
accurate redshifts and the nature of the galaxies have to be confirmed
with additional spectroscopy and/or deep far-infrared imaging with the
{\it Herschel Space Observatory}. 
The ongoing investigation of all 64 high-redshift 3CR sources will result in a
homogeneous database of considerable cosmological impact.


%% Included in this acknowledgments section are examples of the
%% AASTeX hypertext markup commands. Use \url without the optional [HREF]
%% argument when you want to print the url directly in the text. Otherwise,
%% use either \url or \anchor, with the HREF as the first argument and the
%% text to be printed in the second.

\acknowledgments

This work is based in part on observations made with the {\it Spitzer
Space Telescope,}\/ which is operated by the Jet Propulsion Laboratory,
California Institute of Technology under a contract with NASA.
Support for this work was provided by NASA through an award issued by
JPL/Caltech.
The ground-based observations reported here were obtained at the MMT 
Observatory, a
joint facility of the Smithsonian Institution and the University of
Arizona. 
This research has made use of the NASA/IPAC Extragalactic Database (NED) 
%which is operated by the Jet Propulsion
%Laboratory, California Institute of Technology, under contract with
%the National Aeronautics and Space Administration.
and of the Sloan Digital Sky Survey (SDSS DR6).
We thank Dominik Bomans and  Hans Hippelein for valuable 
discussions and comments on the manuscript, and an anonynous 
referee for the constructive, critical report.
M.H. is supported by the 
Nordrhein--Westf\"alische Akademie der Wissenschaften.


%% in parentheses, after the keyword, but they will not be verified.

{\it Facilities:} \facility{Spitzer}, \facility{MMT (MegaCam, SWIRC)},
\facility{Chandra},
\facility{HST}.



\begin{figure}[h!]
\begin{center}
  \includegraphics[width=\columnwidth,angle=0,clip=true]{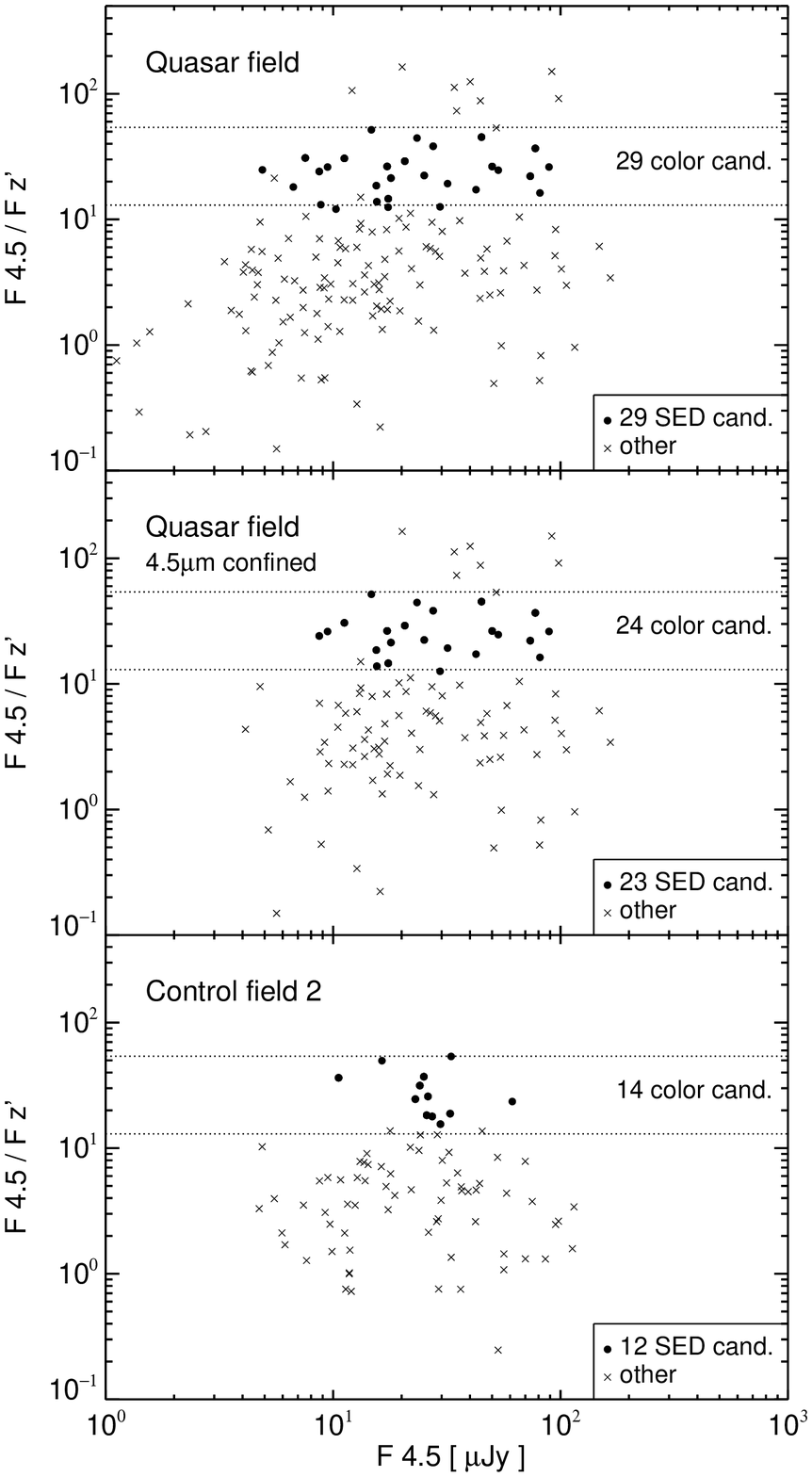}

  \caption{Color-magnitude diagram $F(4.5) / F(z')$ versus  $F(4.5)$
    of the quasar field and control field 2. 
    top: quasar field with all sources detected at 4.5 $\mu$m 
    using SExtractor in double-image mode at 3.6 \& 4.5 $\mu$m. 
    middle: quasar field restricted to those sources detected 
    at 4.5 $\mu$m using SEctractor in single-image mode, as was done
    for the control field 2.  
    bottom: control field 2. 
    Sources with $z_{phot} = 1.53 \pm 0.20$ (SED determined candidates marked with
    circles) concentrate in
    a distinct color range $13 < F(4.5) / F(z') < 54$ indicated by
    the horizontal dotted lines. 
    The number of color determined candidates is also given.
  }
  \label{fig_cm_2}
\end{center}
\end{figure}

\appendix
\section{Source extraction and photometry} 
\label{appendix_a}
For source extraction and photometry, we used the SExtractor tool\footnote{
http://terapix.iap.fr/IMG/pdf/sextractor.pdf}  
(Bertin \& Arnouts 1996, Version 2.5). 
This tool
is able to identify sources via a tree structure algorithm, provide
their total magnitudes,
and, if a saddle point is present in their profile, 
to deblend components via a multi-isophotal analysis. 
Photometry is derived using automatically adjusted apertures
(option MAG-AUTO)
following Kron's first moment algorithm (Kron 1980, Infante 1987).
In case close neighbours are found, the components are 
iteratively deblended, with the 
photometric uncertainty of the faint companion 
reaching 25\% in the worst cases (Beard et al. 1991). 

In a first step we adjusted the SExtractor parameters for each 
image (single-image mode). The parameters are, 
for instance, noise threshold and minimum number of 
connected pixels defining a source. 
The MMT z'- and Y-band images had a seeing of about 0\farcs9
(FWHM), the IRAC PSF has FWHM 
between 1\farcs4 at 3.6~$\mu$m and 1\farcs7 at 8.0~$\mu$m. 
Typical Kron radii of the sources (containing 90\% of the flux) 
are about 3--4 arcsec in all filters (except 24~$\mu$m), so that 
the PSF differences play a minor role ($<5\%$), and  
the $z'$- and $Y$-band photometry --- as computed by SExtractor --- needs no
aperture correction.\footnote{
Our MMT $z'$-band photometry was calibrated with the z-band image of the
Sloan Digital Sky Survey (SDSS, DR6), including a (very small) color
term to account for the similar but nonetheless different bandpasses.
Our $Y$-band photometry was
calibrated with interpolated total magnitudes from SDSS and 2MASS.} 
For the IRAC photometry, however, the absolute photometric calibration 
refers to an aperture of 12\arcsec, so we 
applied standard aperture corrections (5--30\%)\footnote{
http://ssc.spitzer.caltech.edu/irac/dh/}
to derive the final photometry of the sources, all of which are
essentially unresolved. 
The very few sources detected at 24~$\mu$m are well separated, so that
clean photometry could be derived without
need for aperture correction.

In a second step, we ran SExtractor in double-image mode on
the 4.5, 5.8, and 8.0~\micron\ IRAC maps,
using as primary the 3.6 $\mu$m map because it contains more sources
than any other IRAC map.
The double-image mode mode is able to identify, for instance, for a 3.6~$\mu$m
detection also the corresponding 4.5~$\mu$m source, even if it was not
detected in single-image mode. 
% (Sources with z'-band upper limit are not of
% interest for the present study; therefore we did not run 
% SExtractor in double-image mode on the MMT maps.)
The double-image mode enabled us to detect essentially 
all 3.6 $\mu$m sources  at 4.5~$\mu$m (272 of 274 sources). 
The detection rate at 5.8 and 8.0~$\mu$m was 
lower (202 and 212 sources, respectively). We determined the 
5.8 and 8.0~$\mu$m upper flux limits ($\approx$7\,$\mu$Jy) 
from the break of the flux histograms of the detections.
In case of non-detections and clean regions 
we used this flux value as the 3$\sigma$ upper limit.  
% This value corresponds 
% to ${\nu}$F$_{\nu}$ = 3.6$\cdot$10$^{\rm -17}$ 
% W/m$^{\rm 2}$ at 5.8$\mu$m and 2.6$\cdot$10$^{\rm -17}$ 
% W/m$^{\rm 2}$ at 8.0$\mu$m, as shown in Fig. \ref{fig_seds_poor}. 

SExtractor was run separately on the $z'$ image, and the $z'$ and IRAC
source lists were matched with a search radius of 1.5\arcsec.
This should produce good identifications because no two IRAC
sources are closer together than 3\arcsec.
To check further how far confusion plays a role, we counted the number of MMT
sources within a radius of 3\arcsec\ around each IRAC source.
By far most sources ($>$90\%) are unambiguously matched to a single
$z'$ source.  Of 274 IRAC sources in 
the quasar field, 18 (6.6\%) have two
$z'$ counterpart,  and one even has three.  The control fields show
about the same statistics: 
9 double $z'$ counterparts (4.4\%) out of 205 IRAC sources in the
3.6~\micron\ field and 10 double counterparts  (6.0\%) out of 167
IRAC sources in the 4.5~\micron\ field.
Thus, the fraction of confused sources is small and 
comparable for control and quasar fields, and 
therefore we do not expect that the clustering results 
are significantly affected by confusion. 
To avoid any possible complication, we have omitted from the analysis
all sources (4--7\%) 
with two or more $z'$ counterparts found by SExtractor.
The starburst cluster candidate galaxy (object number 3 in 
Table~\ref{table_1}) remained in the sample because SExtractor
treated it as a single source.  Most likely the
saddle between the two components is not deep enough for SExtractor
to split them  despite the source's visual appearance on the $z'$
image as a 0\farcs8 double.

\section{Comparison with the control fields} 
\label{appendix_b}

The quasar field has photometry in six filters, two MMT and four IRAC
bands. But the two control fields have photometry in only three filters, 
in $z'$ and two IRAC bands: 3.6 and 5.8~$\mu$m in control field 1, 
4.5 and 8.0~$\mu$m in control field~2. This makes the uncertainties on the
photometric redshifts larger, and therefore comparison
within the same redshift bin appears not to be straightforward.
Furthermore, in the central field, the source detection rate 
at 4.5~$\mu$m is lower than at 3.6~$\mu$m,   
so that for a fair comparison the number of candidates in control
field~2 has to be corrected for. 
In order to facilitate the comparison, 
we used the colors $z' - [3.6]$ and $z' - [4.5]$ and counted the
number of candidates in the respective color bins. 

Control field 1: 
The colors of the elliptical and starburst template,
when redshifted to $z=1.53$, are $F(3.6)/F(z')$ = 25 and 39, respectively.
In the color range 
$15 < F(3.6)/F(z') < 52$ we found 28 and 17 sources 
for the quasar and the control field, respectively (Fig.~\ref{fig_cm}).
With regard to the low number statistics, these values are 
consistent with the 29 and 16 sources in the quasar and the control
field, having $1.33 < z_{phot} < 1.73$ (denoted as SED candidates in 
Fig.~\ref{fig_cm}). 
This leaves an excess of 11 colour-- and 13 SED--selected sources 
in the quasar field. 

Control field 2: 
The colors of the elliptical and starburst template,
redshifted to $z=1.53$, are $F(4.5)/F(z')$ = 24 and 43, respectively. 
Guided by these ``expectation values'' and the actual $z' - [4.5]$
distribution of the candidates 
in the quasar field, we selected a color range of 
$13 < F(4.5)/F(z') < 54$ 
(Fig.~\ref{fig_cm_2}, top). 
Now we considered the fields, where the 4.5~$\mu$m source list was
created using the 
single-mode SExtractor option, i.e., not making use of the 3.6~$\mu$m
information in the quasar field.
We found 24 and 14 color-selected
candidates for the quasar and the control field, respectively
(Fig.~\ref{fig_cm_2}, middle and bottom). 
These values are consistent with  the corresponding numbers for
SED-selected candidates of 23 and 12 in the quasar and the control
field, respectively. 
Because the quasar field contains more cluster galaxy
candidates, 29 via color and 29 via $z_{phot}$ (Fig.~\ref{fig_cm_2},
top), 
we have scaled up the number of candidates in the control field
by the factors 29/24 for the color-candidates and 29/23 for the
SED-candidates. This results in 17 colour- and 15
SED-selected candidates.
This leaves an excess of 12 ($=29-17$) colour- and 14 ($=29-15$) 
SED-selected sources in the quasar field.

Combining the control field counts we estimate an excess of 11--12
colour- and 13--14 SED-selected candidate cluster galaxies 
in the quasar field.

%\clearpage

\end{document}